\documentclass[12pt,preprint]{aastex}

\shorttitle{Solar convection and oscillations in magnetic regions}
\shortauthors{Jacoutot et al.}

\begin{document}

\title{Realistic numerical simulations of solar convection and oscillations in magnetic regions}
\author{L. Jacoutot$^1$, A. G. Kosovichev$^2$,
A. Wray$^3$, \& N. N. Mansour$^3$}
 \affil{$^1$Center for Turbulence Research, Stanford University, Stanford, CA 94305\\
 $^2$Hansen Experimental Physics Laboratory, Stanford University,
Stanford, CA 94305\\ $^3$NASA Ames Research Center, Moffett field,
CA 94035}

\begin{abstract}
The goal of this research is to investigate how magnetic field affects
the dynamics of granular convection and excitation of
solar oscillations by means of realistic numerical simulations.
We have used a 3D, compressible, non-linear radiative
magnetohydrodynamics code developed at the NASA Ames Research
Center. This code takes into account several physical phenomena:
compressible fluid flow in a highly stratified  medium, sub-grid
scale turbulence models, radiative energy transfer between
the fluid elements, and a real-gas equation of state.
We have studied the influence of the magnetic field of various
strength on the convective cells and on the excitation  mechanisms
of the acoustic oscillations by calculating spectral properties
of the convective motions and oscillations. The results reveal
substantial changes of the granulation structure with increased
magnetic field, and a frequency-dependent reduction in the
oscillation power in a good agreement with solar observations. These simulations
suggest that the enhanced high-frequency acoustic emission at the boundaries of active
region ("acoustic halo" phenomenon) is caused by the changes of the
spatial-temporal spectrum of the turbulent convection in magnetic field,
resulting in turbulent motions of smaller scales and higher frequencies
than in quiet Sun regions.
\end{abstract}


\section{Introduction}
Observations of solar oscillations have revealed that their properties change significantly in magnetic regions \citep{brown1992}. Using SOHO/MDI data \citet{brown1998} found that the power of Doppler-velocity oscillations with frequencies less than 5.2 mHz (which corresponds to the acoustic cut-off frequency) decreases with field strength, but the oscillations at higher frequencies become stronger. The regions of enhanced high-frequency oscillation power are usually observed at the boundaries of active regions and are sometimes referred as "halos". Similar effect was found for local oscillation modes by \citet{howe2004}, who also found that in magnetic regions the line width of acoustic modes increases at low frequencies and decreases at high frequencies. The interpretation of these results is not clear. These variations can be caused by the interaction of acoustic waves with magnetic field and also by changes in the properties of solar convection and excitation mechanism of acoustic waves. It is well-known that magnetic field inhibits convection, thus causing the reduction in the power of acoustic waves produced by the turbulent convective motions. However, the details of this process are unknown. We address this problem by using realistic numerical simulations of solar convection in the presence of magnetic field. The realistic numerical simulations pioneered by \citet{Stein_2001} have provided an important insight into the excitation mechanism of oscillations on the Sun and solar-type stars, showing that the acoustic modes by the work of turbulent pressure (Reynolds stresses) and nonadiabatic gas pressure (entropy) fluctuations \citep{Stein_2004}. Of course, the numerical simulation must rely on sub-grid scale models of turbulence. \citet{jacoutot2008} investigated various turbulence models and showed that the dynamic model of \citet{moin} provides the best agreement with solar observations.
In this paper, we include magnetic field in the realistic simulations and investigate changes in the physical properties of convection and oscillations.

\section{Numerical model}
We use a 3-D, compressible, non-linear radiative-magnetohydrodynamics code
developed by Dr. A. Wray for simulating the upper solar convection zone
and lower atmosphere. This code takes into account several physical
phenomena: compressible fluid flow in a highly stratified medium,
radiative energy transfer between the fluid elements, a real-gas
equation of state, and magnetic effects.
Direct numerical simulation (DNS) of the high Reynolds-number turbulent motions
on the Sun is not achievable. It is not possible to resolve all the motion
scales even with the most advanced computers. The large-eddy
simulation (LES) method allows overcoming the Reynolds-number limits
possible for DNS by modeling the effects of the smallest turbulent
scales. Thus, LES method is used in the present calculations.

The equations we solve are the grid-cell averaged
(henceforth called ``averages'') conservations of mass (\ref{mass}),
momentum (\ref{mom}), energy (\ref{energy}), and magnetic flux (\ref{eqB}):
\begin{equation} \label{mass}
\frac{\partial \rho}{ \partial t}+\left(\rho u_i\right)_{,i}=0,
\end{equation}
\begin{equation} \label{mom}
\frac{\partial \rho u_i}{ \partial t}+\left(\rho u_i u_j +
(P_{ij}+\rho\tau_{ij})\right)_{,j}=-\rho\phi_{,i},
\end{equation}
\begin{equation} \label{energy}
\frac{\partial E}{ \partial t}+\left(E u_i +
(P_{ij}+\rho\tau_{ij})u_j-(\kappa+\kappa_T) T_{,i} +{\left(\frac{c}{4\pi}\right)}^2 \frac{1}{\sigma+\sigma_T}\left(B_{i,j}-B_{j,i} \right)B_j+
F_i^{rad}\right)_{,i}=0,
\end{equation}
\begin{equation} \label{eqB}
\frac{\partial B_i}{ \partial t}+\left( u_j B_i -
u_iB_j -\frac{c^2}{4\pi(\sigma+\sigma_T)} \left(B_{i,j}-B_{j,i} \right) \right)_{,j}=0,
\end{equation}
where $\rho$ is the averaged mass density, $u_i$ is the
Favre-averaged velocity, $B_i$ is the magnetic field, and $E$ is the averaged total energy density
$E=\frac{1}{2}\rho u_i u_i + \rho e + \rho \phi +\frac{1}{8\pi}B_iB_i $, where $\phi$ is
the gravitational potential and $e$ is the Favre-averaged internal
energy density per unit mass. $F_i^{rad}$ is the radiative flux,
which is calculated by solving the radiative transfer equation, and $P_{ij}$
is the averaged stress tensor $P_{ij}=\left(p+{2}\mu
u_{k,k}/{3}+\frac{1}{8\pi}B_kB_k\right)\delta_{ij}-\mu\left(u_{i,j}+u_{j,i}\right)-\frac{1}{4\pi}B_iB_j$,
where $\mu$ is the viscosity. The gas pressure $p$ is a function of
$e$ and $\rho$ through a tabulated equation of state
\citep{Rogers_1996};  $\tau_{ij}$ is the Reynolds stress, $\kappa$
is the molecular thermal conductivity, $\kappa_T$ is the
turbulent thermal conductivity,  $\sigma$
is the molecular electrical conductivity, and $\sigma_T$ is the
turbulent electrical conductivity.

We have carried out simulations using the most widely used subgrid-scale models (SGS): the Smagorinsky model
\citep{smago}. The Reynolds stresses $\tau_{ij}$
are modeled by the usual eddy viscosity assumption using the
large-scale stress tensor $S_{ij}\equiv(u_{i,j}+u_{j,i})/2$:
\begin{equation} \label{tij}
\tau_{ij} - \frac{1}{3}\tau_{kk}\delta_{ij} = - 2 \nu_T \left(
S_{ij} - \frac{1}{3}S_{kk} \delta_{ij} \right).
\end{equation}
The eddy viscosity $\nu_T$ is defined by the Smagorinsky model
\citep{smago}, $\nu_T = C_S \Delta^2 |S|$; and the trace of the
subgrid Reynolds stress $\tau_{kk}$ is modeled using the Yoshizawa's
expression \citep{Yoshizawa}: $\tau_{kk} = 2 C_S \Delta^2 |S|^2$,
where $|S|\equiv \sqrt{2S_{ij}S_{ij}}$ and $\Delta \equiv (\Delta x
\Delta y \Delta z)^{1/3}$, $\Delta x$, $\Delta y$, and $\Delta z$
are the grid step sizes.

The application of the eddy viscosity, $\nu_T$, and the subgrid kinetic
energy, $\tau_{kk}$, results in the compressible Smagorinsky
formulation \citep{moin}:
\begin{equation}
\tau_{ij}=-2C_S\Delta^2|S|(S_{ij}-u_{k,k}\delta_{ij}/3)+2C_C\Delta^2|S|^2\delta_{ij}/3,
\end{equation}
where $C_S$ is the classical Smagorinsky coefficient as used in
incompressible flow modeling; and $C_C$ is a coefficient associated with the
trace of the subgrid Reynolds stress (which is absent in the
incompressible flow formulation). The two parameters must be specified in some
way. For the original Smagorinsky model, constant values are
used. In the present work, we have chosen $C_S=0.2$ and $C_C=0.1$.

The turbulent Prandtl number was taken as unity to set $\kappa_T$.
The molecular viscosity $\mu$ and thermal conductivity $\kappa$ were
neglected as their solar values are exceedingly small.

The turbulent electrical conductivity $\sigma_T$ is calculated by using the extension of the Smagorinsky model to the MHD case \citep{Theobald_1994}, with the flow velocity shear replaced by a shear of the magnetic field (which is the electric current):  $\sigma_T = C_B \Delta^2 |\textbf{j}|$ with $j_i=\epsilon_{ijk}B_{k,j}$ is the resolved electric current; $\epsilon_{ijk}$ is the Levi-Civita symbol; $C_B$ was taken equal to 1. \citep{Theobald_1994}.

In all the following sections, we simulate the upper layers of the convection zone using 66x66x66
grid cells. The region extends 6x6~Mm horizontally and from 5.5~Mm
below the visible surface to 0.5 Mm above the surface. The initial magnetic field is imposed as a vertical uniform component on a snapshot of the preexisting hydrodynamic convection, calculated as described in \citet{jacoutot2008}. The initial values of the field strength are: 100, 300, 600, and 1200 Gauss. The weak field of 100 G has almost no effect. Thus, we present here the three other cases.

\section{Granulation structure}
\begin{figure}
\begin{center}
 \includegraphics[width=1.0\textwidth]{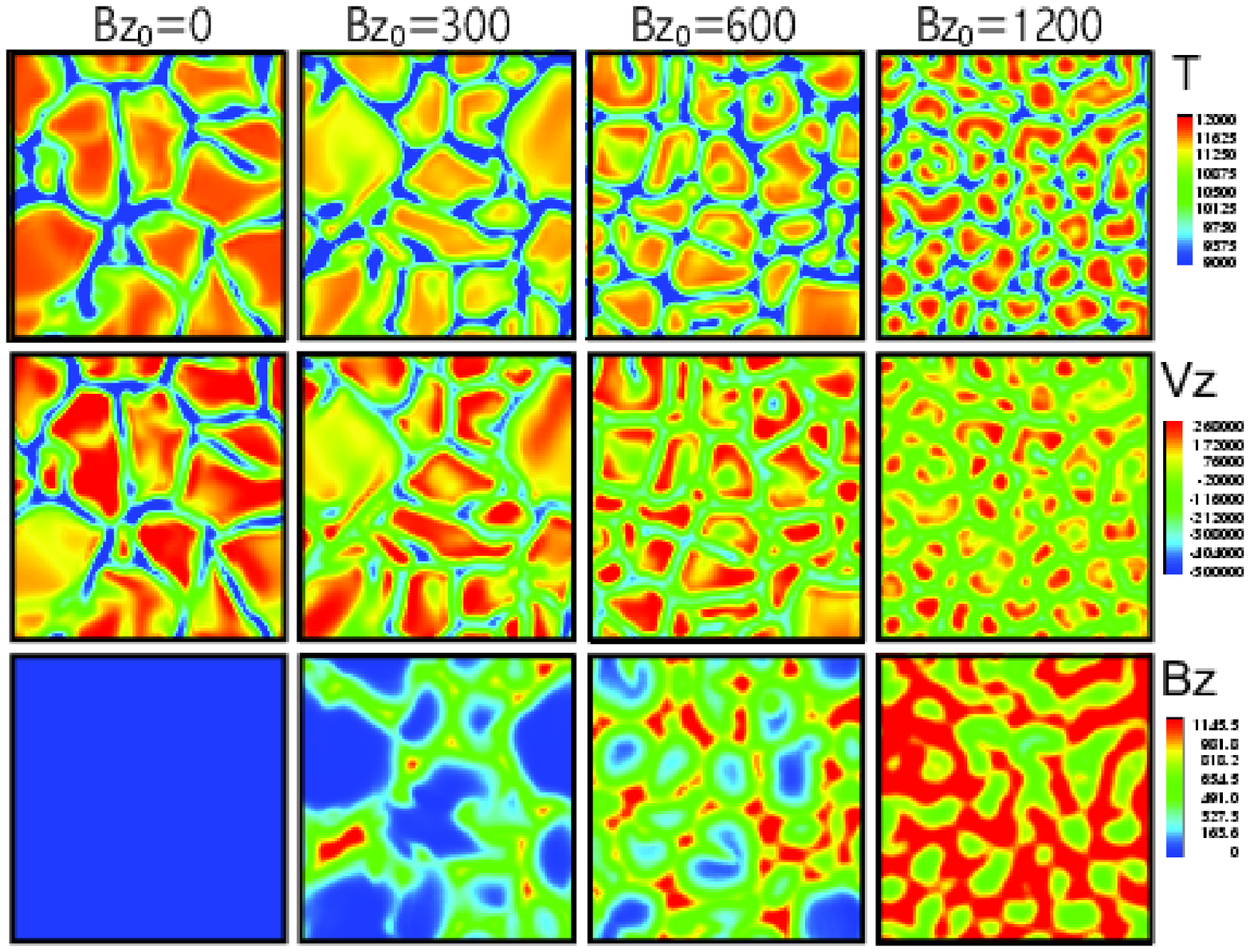}
\caption{Temperature ($K$), vertical velocity ($cm/s$) and vertical magnetic field ($Gauss$) distributions at the visible surface for different initial vertical magnetic fields.
\label{TVzBz}}
\end{center}
\end{figure}

\begin{figure}
\begin{center}
 \includegraphics[width=1.0\textwidth]{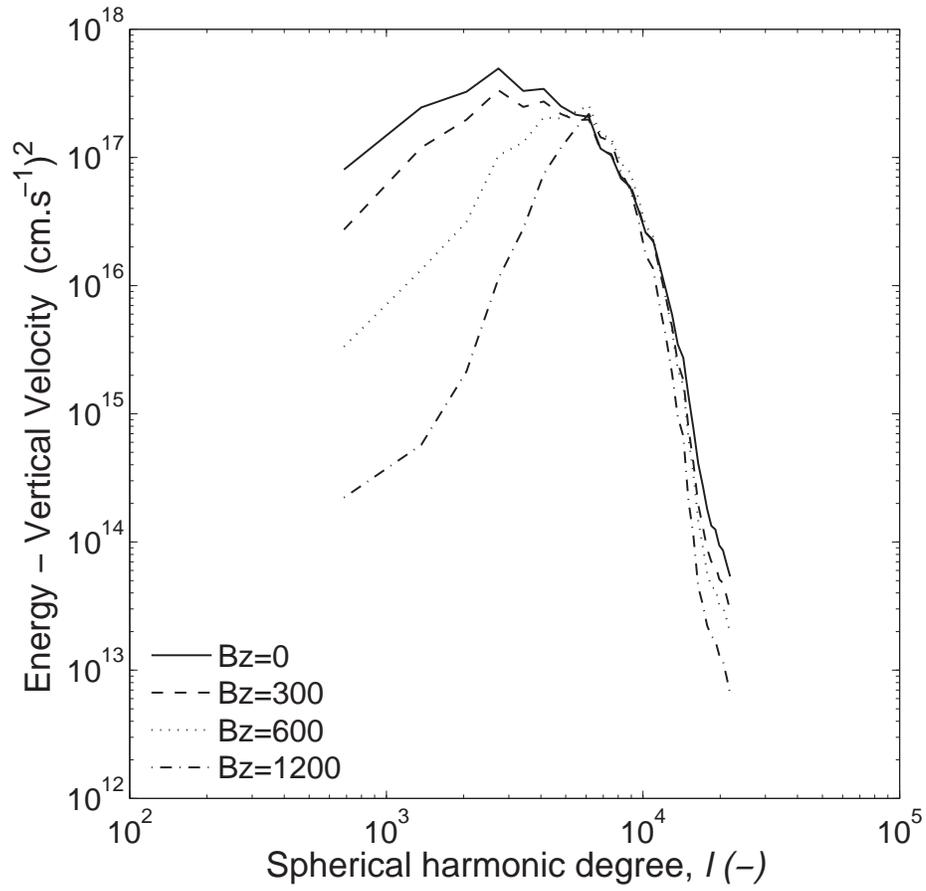}
\caption{Spatial energy of the vertical velocity as a function of the spherical harmonic degree $l$ at the visible surface for different initial vertical magnetic fields.
\label{Ek_w_Bz}}
\end{center}
\end{figure}

The first step was to investigate the influence of the initial magnetic field on the granular structure of convection.
Figure \ref{TVzBz} shows temperature, vertical velocity, and vertical magnetic field distributions at the visible surface for different magnetic fields.
We observe that the size of granules decreases as the initial magnetic field increases. Without magnetic field the mean size of granules is about 2 Mm, and it is less than 0.75 Mm for 1200 Gauss. In addition, the temperature in the granules becomes higher as the initial magnetic field increases. We can also note that the down-flow in the intergranular lanes is weaker for high magnetic fields. Figure \ref{Ek_w_Bz}, which shows the vertical velocity energy as a function of the horizontal wavelength, $k_h$, measured in degrees of spherical harmonics, $l=k_hR_\odot$, demonstrates this feature. The vertical velocity energy is decreased by a factor of $10^3$ for small values $l$, which correspond to the larger granules. We can also notice that the magnetic field is swept into the intergranular lanes although the magnetic field is seeded uniformly. This characteristic of solar magnetoconvection is well-known and has already been presented by \citet{Stein_2002}.

\section{Kinetic energy and oscillation power of radial modes}
\begin{figure}
\begin{center}
 \includegraphics[width=1.0\textwidth]{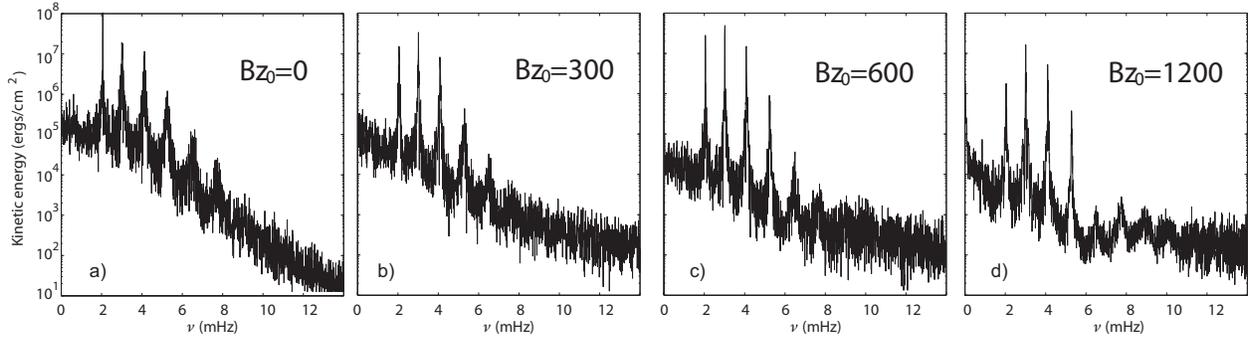}
\caption{Kinetic energy spectra integrated over depth for different initial vertical magnetic fields.
\label{Ew_Bz_4panel}}
\end{center}
\end{figure}

\begin{figure}
\begin{center}
 \includegraphics[width=1.0\textwidth]{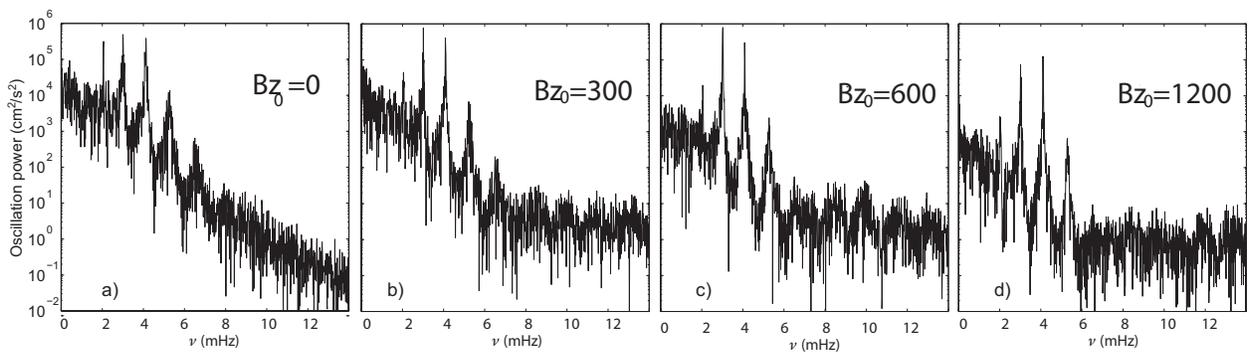}
\caption{Oscillation power spectra at the visible surface for different initial vertical magnetic fields.
\label{P_Bz_4panel}}
\end{center}
\end{figure}

We then studied how the kinetic energy is dissipated for
different initial magnetic field.  To
do this, we calculated the oscillation power spectra of radial
modes. Those modes are extracted by horizontally averaging the
vertical velocity and Fourier-transforming in time.
These results
are obtained from simulations of 60 hours of solar time using
instantaneous snapshots saved every 30 seconds.
Five oscillation modes can be clearly seen as sharp peaks in the spectra of the
horizontally averaged, depth-integrated kinetic energy (Fig.~\ref{Ew_Bz_4panel})
and the power spectrum of the vertical velocity oscillations at the solar surface (Fig.~\ref{P_Bz_4panel}). The
smallest resonant mode frequency is 2.07 mHz. This mode is excited
along all the depth. The resonant frequencies supported by the
computational box are 2.07, 3.03, 4.10, 5.23, 6.52 mHz. In addition, several broad high-frequency peaks at 6-12 mHz corresponding to pseudo-modes \citep{kumar1991} can be identified.

\begin{figure}
\begin{center}
 \includegraphics[width=0.5\textwidth]{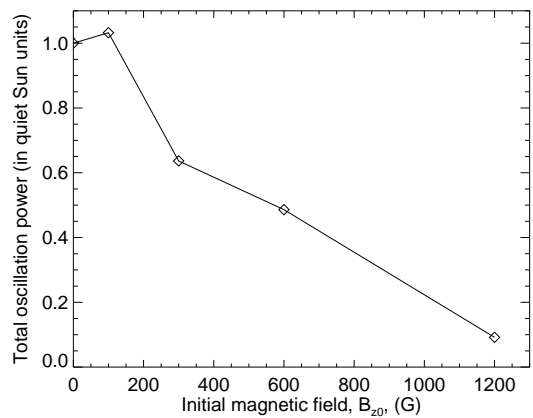}
\caption{Total oscillation power at the visible surface as a function of initial
magnetic field strength.
\label{Total_P__Binit}}
\end{center}
\end{figure}

It is clear that the magnetic field significantly affects the kinetic energy spectrum. The amplitude of the excited modes decreases as the initial magnetic field increases,
as well as the total oscillation power (Fig.~\ref{Total_P__Binit}).
The power distribution is shifted towards higher frequencies with the increase of the field strength. It is particularly interesting that the amplitude of the pseudo-modes increases with field strength and reaches maximum at $B_{z0}=600$ G. This may explain the effect of enhanced high-frequency emission ("acoustic halos") around active regions \citep{brown1992,brown1998,jain2002}. The enhanced emission at frequencies 5-7 mHz appears at the boundaries of active regions where magnetic field is moderate. This corresponds to the simulations results: the pseudo-mode amplitude is high for 600 G field and diminishes at 1200 G.

The simulations also show the enhanced spectral power of the convective background at high frequencies for models with magnetic field, forming plateaux at $\nu > 6$ mHz. This leads to the idea that the acoustic halos are caused by enhanced high-frequency turbulent convective motions in the presence of moderate magnetic field. This is consistent with the decreased granular size in magnetic regions, described in the previous section. The smaller scale convection naturally has higher frequencies and, thus, generates more higher frequency acoustic waves than convection without magnetic field. When the field is very strong the sound generation decreases because of suppression of convective motions of all scales. This probably explains why the acoustic halos are observed in regions of moderate magnetic field strength at the boundaries of active regions.

In addition, the simulation results show that the modal lines in the oscillation
power spectrum (Fig.~\ref{P_Bz_4panel}) becomes broader for the mode of 2.07 and 3.03 mHz (the first mode almost disappears at $B_{z0}=1200$ G), but they are more narrow for the modes of 4.10, 5.23, 6.52 mHz. This qualitatively corresponds to the observational result of \citet{howe2004}. We plan to present a quantitative analysis in a future publication.


\section{Calculation of the p-mode work integrand}

The dominant driving of the p-modes comes from the interaction of the non-adiabatic, incoherent pressure fluctuations with the coherent mode displacement \citep{Nordlund2001}.
Previous work of \citet{Balmforth_1992}, \citet{Goldreich_1994}, \citet{Kumar_1994}, \citet{Nordlund_1998} showed that turbulent motions stochastically excite the resonant modes via Reynolds stresses (turbulent pressure) and entropy fluctuations (gas pressure).
Many analytical formulations have been proposed by \citet{Balmforth_1992, Goldreich_1994, Samadi_2001}. The problem is that these models introduce free parameters related to the choice of the turbulent medium model. The advantage of 3-D realistic numerical simulations comes from the possibility of calculating all the quantities related to turbulent convection. \citet{Nordlund2001} presented a formalism for analyzing the interaction of convection with purely radial oscillations. This formalism seems to be the most accurate because it accounts for phase relations between pressure fluctuations (both turbulent and
gas) and the mode compression factor $\partial \xi_{\omega} /
\partial r$ \citep{Stein_2004}. In this section the mode excitation
rate is calculated using the same method, but we add the contribution of the magnetic pressure $\frac{1}{8\pi}B_kB_k$ in the calculation of the total pressure. The rate of energy input
to the modes per unit surface area ($erg.cm^{-2}.s^{-1}$) is
\begin{equation}
\label{eqn:rate} \frac{\Delta <E_{\omega}>}{\Delta t}=\frac{\omega^2
|\int_r dr \delta P_{\omega}^*  \left( \partial \xi_{\omega} /
\partial r \right)|^2}{8\Delta \nu E_{\omega}},
\end{equation}
where $\delta P_{\omega}^*$ is the Fourier transform of the
non-adiabatic total pressure; $\delta$ in front of the pressure
means that one computes so-called 'pseudo-Lagrangian' fluctuations
relative to a fixed mass radial coordinate system
\citep{Nordlund2001}; $\Delta \nu$ is the frequency interval for the
Fourier transform; $\xi_{\omega}$ is the mode displacement for the
radial mode of angular frequency $\omega$. It is obtained from the
eigenmode calculations for a standard solar model of
\citet{Chrsit1996}. His spherically symmetric model S gives 35
radial modes that provide much denser frequency spectrum in
comparison with the three resonant modes obtained within the
simulation box. $E_{\omega}$ is the mode energy per unit surface
area ($erg.cm^{-2}$) defined as:
\begin{equation}
\label{eqn:Ew} E_{\omega}=\frac{1}{2} \omega^2 \int_r dr \rho
{\xi_{\omega}}^2 {\left(\frac{r}{R} \right)}^2.
\end{equation}

\begin{figure}
\begin{center}
 \includegraphics[width=1.0\textwidth]{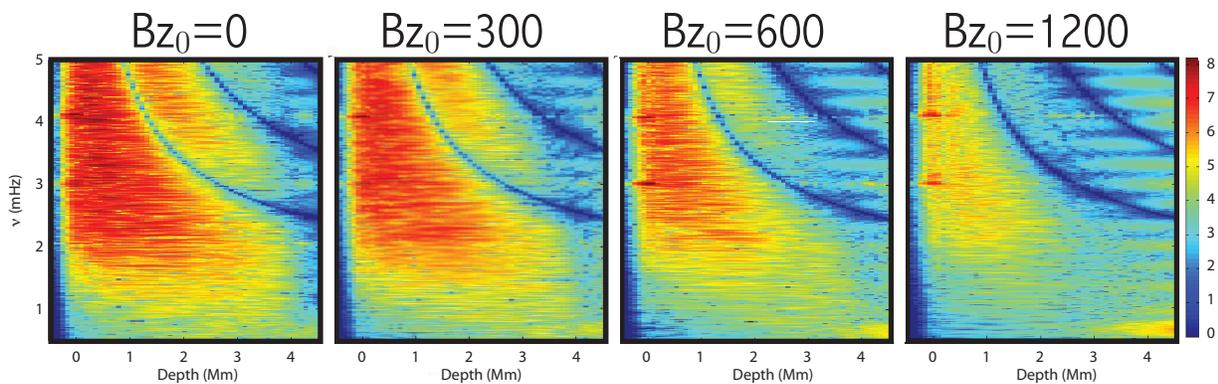}
\caption{Logarithm of the total work integrand (eq. (\ref{eqn:rate}) in
units of $erg\,cm^{-2}\,s^{-1}$), as a function of depth and
frequency for different initial vertical magnetic field.
\label{Ew}}
\end{center}
\end{figure}

\begin{figure}
\begin{center}
 \includegraphics[width=1.0\textwidth]{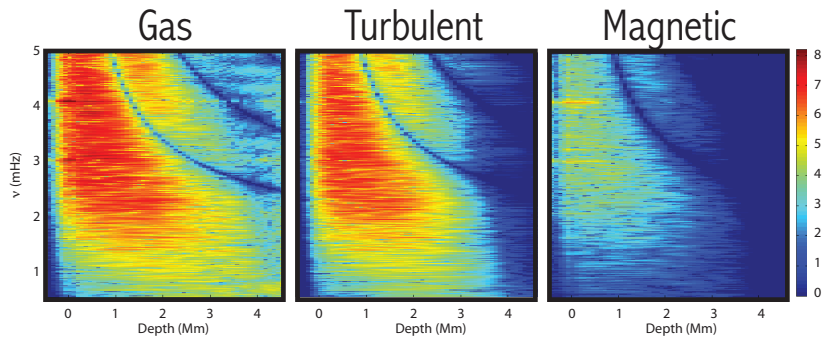}
\caption{Gas, turbulent and magnetic pressure contributions in the work integrand (in
units of $erg\,cm^{-2}\,s^{-1}$), as a function of depth and
frequency for an initial vertical magnetic field of 300~Gauss.
\label{Ew_gtet}}
\end{center}
\end{figure}

The distributions of the integrand of the work integral as a
function of depth and frequency (Figure \ref{Ew}) are similar to the results obtained by \citet{Stein_2001}. Most driving
is concentrated between the surface and 500 km depth at around 3-4
mHz. We can see that the excitation becomes weaker for high initial magnetic field. The contribution of the different pressures (gas, turbulent and magnetic)
 is presented in Fig.~\ref{Ew_gtet}. We see that the dominant driving comes from the  interaction of the coherent mode displacement with the non-adiabatic, incoherent gas pressure fluctuations. The contribution of the turbulent pressure shows similar shape than that of the gas pressure, but it is weaker. The magnetic pressure plays a weak role in the excitation of the p-modes for 300 Gauss. This contribution becomes higher with increased magnetic field. The total work integrand with increased magnetic field becomes localized closer to the surface and at the mode frequencies.

\section{Conclusion}
We have carried out realistic simulations of solar convection and oscillation in the presence of magnetic field of various strength. Initially, the magnetic field
was vertical and uniformly distributed. The results reproduced several phenomena observed in solar magnetic regions.

In particular, the results confirm that the spatial scale of granulation substantially decreases with the magnetic field strength. Magnetic field is swept in the intergranular lanes, and the vertical downdraft motions in these lanes are suppressed. This results is a decrease in the excitation power. The oscillation power in the presence of magnetic field is shifted towards higher frequencies, also increasing the amplitude of pseudo-modes above the acoustic cut-off frequency. At a moderate field strength of $\sim 600$ G the power of the high-frequency oscillations reaches a maximum. This corresponds to the phenomenon of "acoustic halo" observed in the range of 5-7 mHz at the boundaries of active regions. The reason of this phenomenon is probably in the change of the spatial-temporal spectrum of the turbulent convection in magnetic field. The convective motions becomes smaller in spatial scale and faster in the presence of magnetic field, and this causes the changes in the oscillations excited by these motions.

The power spectra show a significant increase of the convective power ("plateau") at high frequencies. The oscillation modal lines becomes broader and smaller in amplitude for the low-frequencies modes, but more narrow at higher frequencies. Qualitatively, this corresponds to the observations, but requires a more detailed quantitative study.

The calculations of the work integrand show that the excitation mechanism in magnetic regions remains the same as in the quiet Sun. The oscillations are excited by fluctuations of Reynolds stresses and entropy with magnetic forces playing a minor role. With increased magnetic field the work integrand becomes more concentrated in the near-surface layers and at the mode frequencies.

In general, our simulations lead to a conclusion that the observed changes of solar oscillations in magnetic regions on the Sun are mostly caused by changes in the spatial-temporal spectrum of convective motions, which are the source of the oscillations.

\clearpage

\clearpage

\end{document}